\documentclass[conference,a4paper]{IEEEtran}
\IEEEoverridecommandlockouts
\usepackage{color}
\usepackage{graphicx}
\usepackage{epstopdf}
\usepackage{amsmath}
\usepackage{amssymb} 
\usepackage{algorithm}
\usepackage{algorithmic}
\usepackage{amsmath}
\usepackage{multirow}
\usepackage{booktabs}
\usepackage{array}
\usepackage{amsthm}
\usepackage{stfloats}
\usepackage{caption}
\usepackage{subfigure}
\usepackage{bm}
\usepackage{setspace}
\usepackage{diagbox}
\usepackage{flushend}
\usepackage{comment}

\newcommand{\be}{\begin{equation}}
\newcommand{\ee}{\end{equation}}
\newcommand{\bea}{\begin{eqnarray}}
\newcommand{\eea}{\end{eqnarray}}
\newcommand{\ba}{\begin{array}}
\newcommand{\ea}{\end{array}}

\def\BibTeX{{\rm B\kern-.05em{\sc i\kern-.025em b}\kern-.08em
    T\kern-.1667em\lower.7ex\hbox{E}\kern-.125emX}}
\begin{document}

\title{Mutual-Coupling-Aware Electromagnetic and Signal Processing Co-Design for Pixel Antenna Arrays
}

\author{ \IEEEauthorblockN{Ping Ma$^{\dag}$,  Mengzhen Liu$^{\dag}$, Rang Liu$^{\ddag}$, Ming Li$^{\dagger*}$, and A. Lee Swindlehurst$^{\S}$\\
		\vspace{-0.3 cm} }
	\IEEEauthorblockA{$^{\dag}$Dalian University of Technology, Dalian, Liaoning 116024, China\\
    	E-mail: \texttt{\{pingma, liumengzhen\}@mail.dlut.edu.cn, mli@dlut.edu.cn}}

        \IEEEauthorblockA{$^{\ddag}$Friedrich-Alexander-University Erlangen-Nuremberg (FAU), Erlangen 91058, Germany, E-mail: \texttt{rang.liu@fau.de }}
        \IEEEauthorblockA{$^{\S}$University of California, Irvine, CA 92697, USA, E-mail: \texttt{swindle@uci.edu }}
        \vspace{-0.5 cm}	}
	
\maketitle

\pagestyle{empty}  % no page number for the second and the later pages
\thispagestyle{empty} % no page number for the first page
% \vspace{-0.3 cm}

\begin{abstract}

Reconfigurable pixel antenna arrays provide additional electromagnetic (EM) degrees of freedom for wireless transmission by enabling hardware-level adaptation of radiation responses. In practical arrays, inter-element mutual coupling (MC) alters the switch-state-dependent responses, causing MC-unaware designs to be mismatched with the realized coupled-array channels. To address this issue, we propose an MC-aware EM and signal-processing co-design framework for downlink multi-user multiple-input single-output (MU-MISO) systems. The full pixel antenna array is modeled as a unified multi-port circuit network that jointly characterizes the feed ports and switch-controlled pixel ports. By combining the impedance matrix extracted from Computer Simulation Technology (CST) Studio Suite simulations with the full-array open-circuit far-field responses, we construct a switch-state-dependent EM-domain channel that explicitly captures both intra-antenna pixel coupling and inter-element MC. Based on this model, we formulate a sum-rate maximization problem over the binary switch-state vector and the transmit precoder. To address the resulting mixed-integer non-convex problem, we develop an alternating optimization algorithm in which the binary switch-state vector is updated via sequential exhaustive block optimization (SEBO), while the transmit precoder is optimized using fractional programming (FP). CST-based simulations show that the proposed MC-aware design outperforms MC-unaware, MC-free, and conventional patch-array baselines. These results demonstrate that accurate MC modeling is essential not only for reducing model mismatch, but also for exploiting the coupled embedded responses of practical pixel antenna arrays.

\end{abstract}


\section{Introduction}
Reconfigurable antennas introduce electromagnetic (EM) degrees of freedom beyond those available in conventional fixed antenna arrays~\cite{M.Liu_Reconfigurable,S. Shen_Reconfigurable}. 
Among various reconfigurable antenna architectures, pixel antenna arrays are particularly attractive because they discretize a radiating surface into switch-controlled metallic pixels, enabling flexible reshaping of the surface current distribution and the corresponding radiation response through adaptive pixel interconnections ~\cite{A.Grau_Multifunctional}-\cite{S.Song_Efficient}. 
Such hardware-level radiation reconfigurability enables the
transmitter to adapt not only the baseband precoder, but also the
underlying antenna radiation behavior, thereby opening new
opportunities for EM and signal processing co-design
in wireless communication systems.

The performance of such co-design approaches depends heavily on the accuracy of the EM model used for communication-oriented optimization. These models must accurately account for the electromagnetic coupling between closely spaced antennas, where currents induced on one antenna element excite nearby elements, leading to interactions among their radiated EM fields. This inter-element mutual coupling (MC) can substantially distort the actual radiation response and the resulting effective communication channels~\cite{X.Chen_Mutual}, \cite{H.Yang_Mutual}. In pixel antenna arrays, this issue is even more pronounced because the selected switch states affect not only the intra-element current distribution, but also the MC behavior of the full array. Consequently, antenna configurations designed under idealized MC-unaware assumptions may suffer from beam mismatch and rate loss when applied to coupled arrays.

Recent studies have investigated reconfigurable pixel-based
antennas for improving wireless communication performance
through joint EM and signal processing design~\cite{S. Shen_Reconfigurable}-\cite{B.Zuo_Antenna}. 
Multi-port-circuit-based formulations are especially useful for relating antenna configurations, port currents, and radiated fields in a unified framework~\cite{H.Li_MultiUser}. However, most existing work either relies on idealized radiation models, focuses mainly on isolated antenna elements, or considers only the controllable interactions within individual pixel antennas \cite{A.Sams_Optimizing}. The unintended inter-element coupling in practical arrays is often ignored or only implicitly approximated. This modeling gap becomes particularly critical in multi-user multiple-input single-output (MU-MISO) systems, where a common antenna configuration simultaneously shapes the effective channels of all users and the resulting inter-user interference. Therefore, an accurate MC-aware array model is essential for exploiting pixel antenna reconfigurability in multi-user systems. Different from approaches that characterize individual antenna elements and subsequently construct the array response, we perform full-wave EM simulation of the entire pixel antenna array in CST and directly extract the impedance matrix and open-circuit embedded radiation responses from the coupled array. As a result, the extracted EM parameters inherently contain both the interactions among the constituent pixels and the inter-element mutual coupling among different pixel antennas. These full-array EM parameters are then incorporated into the multi-port circuit model, enabling the switch-state-dependent radiation response and communication channel to explicitly reflect the actual coupled-array behavior.

Motivated by these observations, we develop an 
MC-aware EM and signal processing co-design
framework for a downlink MU-MISO system, where a base station (BS) equipped with a pixel antenna array serves multiple single-antenna users. The entire array is modeled as a unified multi-port circuit network, in which the feed ports and switch-controlled pixel ports are jointly characterized. Using the impedance matrix and open-circuit radiation responses extracted from Computer Simulation Technology (CST) Studio Suite simulations \cite{benitzhak_cst}, the effects of antenna configuration and inter-element MC are explicitly embedded into the EM-domain channel model. Based on this
channel model, we formulate a sum-rate maximization problem over the binary antenna switch-state and the transmit precoder. To tackle the resulting mixed-integer non-convex problem, we develop an alternating optimization algorithm in which the switch-state vector is updated via sequential exhaustive block optimization (SEBO), while the transmit precoder is optimized using fractional programming (FP) to manage multi-user interference under the total transmit-power constraint. Simulation results using CST-extracted array responses show that the proposed MC-aware design outperforms MC-unaware, MC-free, and conventional patch antenna baselines, demonstrating the
importance of explicitly accounting for MC in EM and signal-processing co-design for practical pixel antenna arrays.

\begin{figure}[!t]
 \centering
 \includegraphics[width= 0.7\linewidth]{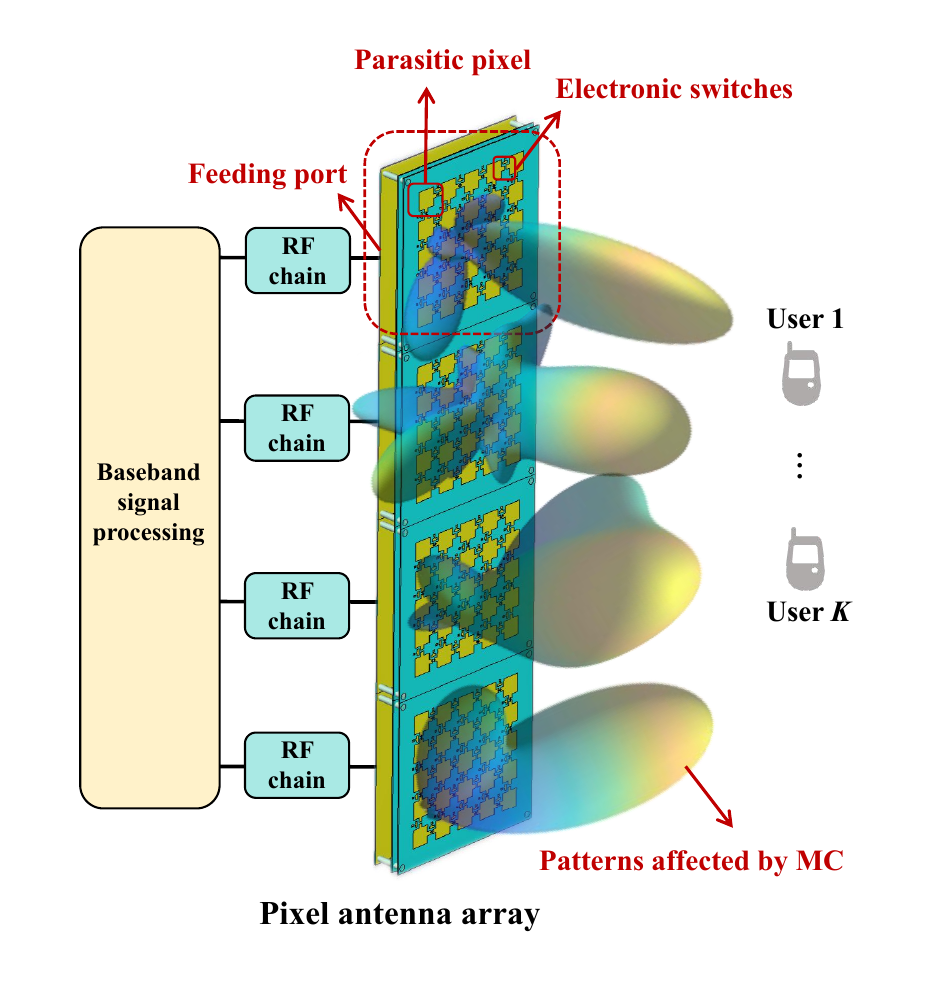}  \vspace{-0.5 cm}
 \caption{A downlink MU-MISO communication system with pixel antenna arrays.}
 \label{fig:systemmodel}
  \vspace{-0.4 cm}
\end{figure}

\section{Pixel Antenna Array}

\setcounter{equation}{1}
As illustrated in Fig. 1, each pixel antenna is implemented by discretizing a radiating
surface into parasitic metallic pixels. Adjacent pixels are interconnected through electronic switches, whose on/off states enable dynamic configuration of the antenna topology. Through this mechanism, the EM coupling among constituent pixels can be intentionally controlled, thereby modifying the current distribution and reshaping the radiated EM fields. 
Manipulation of the radiation of the pixel antennas leads to current variations in each antenna that inevitably induce EM interactions with other elements. This effect introduces uncontrolled inter-element MC among the antennas, which alters the realized embedded vector radiation
responses in both the angle and polarization domains. Without
accurate MC modeling, this leads to channel mismatch and rate loss. 

Compared with conventional patch antennas with fixed radiation characteristics, pixel antennas provide additional EM-domain degrees of freedom by reconfiguring the surface current distribution through switch-controlled pixel interconnections. This enables adaptive radiation-pattern and polarization responses without increasing the number of RF chains, which is particularly beneficial for multi-user transmission where the effective channels can be shaped according to the user directions. However, such flexibility comes at the cost of additional switching and biasing circuitry, and more complicated EM interactions among both pixels and neighboring antenna elements. Therefore, pixel antennas generally require more sophisticated configuration optimization and more accurate EM modeling than conventional patch antennas.  

\begin{figure}[!t]
 \centering
 \includegraphics[width= 0.7\linewidth]{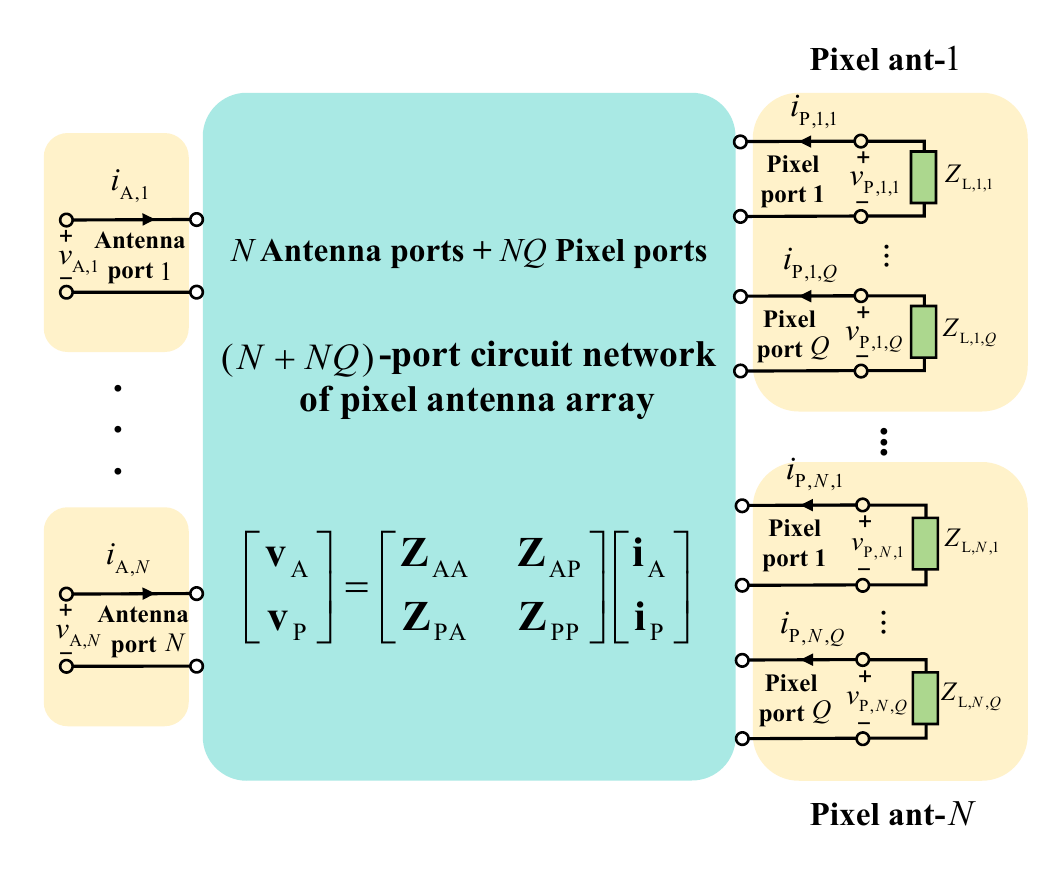}  \vspace{-0.4cm}
 \caption{Schematic of the multi-port circuit network for a pixel antenna array.}
 \label{fig:pixel_antenna}
  \vspace{-0.4 cm}
\end{figure}

In this work, we apply multi-port  circuit network theory to establish a unified modeling framework for all pixel antennas within the array, enabling a natural embedding of the MC effects in the array response. In particular, as depicted in the schematic of Fig.~\ref{fig:pixel_antenna}, the pixel antenna array is modeled as an \((N + N Q)\)-port circuit network that consists of \(N\) antenna ports for the signal feeds and \(Q\) pixel ports for the connections between adjacent pixels in each antenna. The EM interactions among these ports are captured by induced voltages and currents, whose relationships are characterized by an impedance matrix,  as expressed in \eqref{eq:expanded_Z} shown at the top of next page. In this expression, \(v_{\mathrm{A},n}\in \mathbb{C}\) and \(\mathbf{v}_{\mathrm{P},n}\in \mathbb{C}^{Q}\) represent the voltages at the antenna and pixel ports of the $n$-th antenna, respectively. The currents \(i_{\mathrm{A},n}\in \mathbb{C}\) and \(\mathbf{i}_{\mathrm{P},n}\in \mathbb{C}^{Q}\) follow the same definition. The impedance entries
\(z_{\mathrm{AA},mn}\in \mathbb{C}\) and \(\mathbf{Z}_{\mathrm{PP},mn} \in \mathbb{C}^{Q \times Q}\) represent the self impedance (for $m=n$) or transfer impedance (for $m\neq n$) associated with the antenna and pixel ports, respectively. The components  \(\mathbf{z}_{\mathrm{PA},mn} \in \mathbb{C}^{Q}\) and \(\mathbf{z}_{\mathrm{AP},mn} \in \mathbb{C}^{Q}\) denote the transfer impedances between the antenna and the pixel ports for the $m$-th and $n$-th antenna elements. To facilitate the analysis, we further describe the network equation in \eqref{eq:expanded_Z} in the following block-wise form: 
\begin{equation}\label{eq:Z_matrix}
\left[
  \begin{array}{c}
    \mathbf{v}_{\mathrm{A}}  \\
    \mathbf{v}_{\mathrm{P}}
  \end{array}
 \right]=
\left[
  \begin{array}{cc}
    \mathbf{Z}_{\mathrm{AA}} & \mathbf{Z}_{\mathrm{AP}}  \\
    \mathbf{Z}_{\mathrm{PA}} & \mathbf{Z}_{\mathrm{PP}}
  \end{array}
 \right]
 \left[
  \begin{array}{c}
    \mathbf{i}_{\mathrm{A}}  \\
    \mathbf{i}_{\mathrm{P}}
  \end{array}
 \right],
\end{equation}
with  the voltages denoted as \(\mathbf{v}_{\mathrm{A}}\triangleq[v_{\mathrm{A},1},v_{\mathrm{A},2}, \ldots, v_{\mathrm{A},N}]^{T}\in \mathbb{C}^{N}\), \(\mathbf{v}_{\mathrm{P}}\triangleq[\mathbf{v}_{\mathrm{P},1}^{T},\mathbf{v}_{\mathrm{P},2}^{T}, \ldots, \mathbf{v}_{\mathrm{P},N}^{T}]^{T}\in \mathbb{C}^{NQ}\), and the currents as \(\mathbf{i}_{\mathrm{A}}\triangleq[i_{\mathrm{A},1},i_{\mathrm{A},2}, \ldots, i_{\mathrm{A},N}]^{T}\in \mathbb{C}^{N}\), \(\mathbf{i}_{\mathrm{P}}\triangleq[\mathbf{i}_{\mathrm{P},1}^{T},\mathbf{i}_{\mathrm{P},2}^{T}, \ldots, \mathbf{i}_{\mathrm{P},N}^{T}]^{T}\in \mathbb{C}^{NQ}\). 

\begin{figure*}[!t]
 \setcounter{equation}{0}
 \small
\begin{equation}\label{eq:expanded_Z}
\left[
  \begin{array}{c}
    v_{\mathrm{A},1} \\
    v_{\mathrm{A},2} \\
   \vdots \\
    v_{\mathrm{A},N} \\[6pt]
     \hline
     \rule{0pt}{16pt}
     \mathbf{v}_{\mathrm{P},1} \\
     \mathbf{v}_{\mathrm{P},2} \\
     \vdots \\
     \mathbf{v}_{\mathrm{P},N} \\
  \end{array}
\right]=
\left[
  \begin{array}{cccc|cccc}
    z_{\mathrm{AA},11}& z_{\mathrm{AA},12}&\ldots& z_{\mathrm{AA},1N}&\mathbf{z}_{\mathrm{AP},11}^{T} &\mathbf{z}_{\mathrm{AP},12}^{T}&\ldots&\mathbf{z}_{\mathrm{AP},1N}^{T}\\
      z_{\mathrm{AA},21}&z_{\mathrm{AA},22}&\ldots& z_{\mathrm{AA},2N}&\mathbf{z}_{\mathrm{AP},21}^{T} &\mathbf{z}_{\mathrm{AP},22}^{T}&\ldots&\mathbf{z}_{\mathrm{AP},2N}^{T}\\
    \vdots&\vdots&\ddots& \vdots &\vdots &\vdots&\ddots&\vdots\\
     z_{\mathrm{AA},N1}&z_{\mathrm{AA},N2}&&z_{\mathrm{AA},NN}&\mathbf{z}_{\mathrm{AP},N1}^{T} &\mathbf{z}_{\mathrm{AP},N2}^{T}&\ldots&\mathbf{z}_{\mathrm{AP},NN}^{T}\\[6pt]
     \hline
     \rule{0pt}{16pt}
     \mathbf{z}_{\mathrm{PA},11} &\mathbf{z}_{\mathrm{PA},12}&\ldots&\mathbf{z}_{\mathrm{PA},1N}& \mathbf{Z}_{\mathrm{PP},11} &\mathbf{Z}_{\mathrm{PP},12}&\ldots&\mathbf{Z}_{\mathrm{PP},1N} \\
     \mathbf{z}_{\mathrm{PA},21} &\mathbf{z}_{\mathrm{PA},22}&\ldots&\mathbf{z}_{\mathrm{PA},2N}&\mathbf{Z}_{\mathrm{PP},21} &\mathbf{Z}_{\mathrm{PP},22}&\ldots&\mathbf{Z}_{\mathrm{PP},2N}\\
     \vdots &\vdots&\ddots&\vdots&\vdots &\vdots&\ddots&\vdots\\
     \mathbf{z}_{\mathrm{PA},N1} &\mathbf{z}_{\mathrm{PA},N2}&\ldots&\mathbf{z}_{\mathrm{PA},NN}&\mathbf{Z}_{\mathrm{PP},N1} &\mathbf{Z}_{\mathrm{PP},N2}&\ldots&\mathbf{Z}_{\mathrm{PP},NN}\\
  \end{array}
\right]
\left[
  \begin{array}{c}
    i_{\mathrm{A},1} \\
    i_{\mathrm{A},2} \\
   \vdots \\
    i_{\mathrm{A},N} \\[6pt]
     \hline
     \rule{0pt}{16pt}
     \mathbf{i}_{\mathrm{P},1} \\
     \mathbf{i}_{\mathrm{P},2} \\
     \vdots \\
     \mathbf{i}_{\mathrm{P},N} \\
  \end{array}
\right].
\end{equation}
 \hrulefill
\vspace{-0.2 cm}
\end{figure*}

\setcounter{equation}{2}
To capture the reconfigurable switching states of the pixel antennas, we introduce a binary vector 
\begin{equation}
\begin{aligned}\label{eq:b}
 \mathbf{b} &\triangleq [b_{1,1},b_{1,2},\ldots,b_{1,Q},\ldots,  \\
 &\quad \quad \quad \quad b_{N,1},b_{N,2},\ldots, b_{N,Q}]^{T}\in\{0,1\}^{NQ},
 \end{aligned}
\end{equation}
which indicates the states  of the pixel ports  for the $N$ antennas. Each switch state determines the corresponding  load impedance of the corresponding pixel port. Specifically, a switch configured in the OFF state ($b_{n,q}=0$) results in an open-circuit port with impedance \(Z_{\mathrm{off}}\), whereas a switch in the ON state ($b_{n,q}=1$) leads to a short-circuit port with impedance \(Z_{\mathrm{on}}\), satisfying \(|Z_{\mathrm{off}}|\gg |Z_{\mathrm{on}}|\). The ideal open/short switch model is recovered as the limiting case \( |Z_{\mathrm{off}}|\to\infty \) and \(Z_{\mathrm{on}}\to 0\), but the matrix operations below use finite values. This relationship can be summarized as follows
\begin{equation}\label{eq:Z_Lq}
\begin{aligned}
Z_{n,q} =
\begin{cases}
Z_{\mathrm{off}}, & \text{if } b_{n,q} =0, \\
Z_{\mathrm{on}}, & \text{if } b_{n,q} = 1.
\end{cases}
\end{aligned}
\end{equation}
According to the voltage–current relationship of the pixel ports, we can derive the following equation
\begin{equation}\label{eq:load_equation}
\mathbf{v}_{\mathrm{P}} = -\mathbf{Z}(\mathbf{b})\mathbf{i}_{\mathrm{P}},
\end{equation}
where $\mathbf{Z}(\mathbf{b}) \triangleq \mathrm{diag}(Z_{1,1}, Z_{1,2}, \ldots, Z_{N,Q}) \in \mathbb{C}^{NQ \times NQ} $ is the diagonal load impedance matrix.  
Combining \eqref{eq:Z_matrix} and \eqref{eq:load_equation}, the currents at the pixel ports can be expressed as functions of the feed currents at the antenna ports:
\begin{equation}\label{eq:pixel_current}
\mathbf{i}_{\mathrm{P}}(\mathbf{b}) = -(\mathbf{Z}_{\mathrm{PP}}+\mathbf{Z}(\mathbf{b}))^{-1}\mathbf{Z}_{\mathrm{PA}}\mathbf{i}_{\mathrm{A}}.
\end{equation}
From this expression, it can be observed that the currents at the pixel ports are influenced not only by the currents of their associated antennas, but also by those of other antennas in the array, reflecting the presence of MC.

As discussed above, controlling the binary switch-state vector $\mathbf{b}$ alters the pixel port currents, which in turn modifies the radiation patterns and polarization states of the pixel antennas. Due to MC effects, the radiation response of each antenna element depends not only on its own currents but also on those induced in other elements. In what follows, we further analyze the impact of the resulting coupled current distribution on the resulting radiation responses. 
The overall radiation response of the pixel antenna array can be expressed as a superposition of open-circuit radiation response bases, each corresponding to the response when a unit current is fed into a single port with all other ports open-circuited. Using this approach, the radiation response for the entire pixel antenna array can be constructed as
\begin{equation}\label{eq:pattern_n}
\begin{aligned}
\mathbf{e}(\mathbf{b})
&= \mathbf{E}_{\mathrm{A}}^{\mathrm{oc}}\mathbf{i}_{\mathrm{A}}
 + \mathbf{E}_{\mathrm{P}}^{\mathrm{oc}}\mathbf{i}_{\mathrm{P}}(\mathbf{b}) \\
&= \left[\mathbf{E}_{\mathrm{A}}^{\mathrm{oc}}
-\mathbf{E}_{\mathrm{P}}^{\mathrm{oc}}
\left(\mathbf{Z}_{\mathrm{PP}}+\mathbf{Z}(\mathbf{b})\right)^{-1}
\mathbf{Z}_{\mathrm{PA}}\right]\mathbf{i}_{\mathrm{A}},
\end{aligned}
\end{equation}
where  \(\mathbf{E}_{\mathrm{A}}^{\mathrm{oc}}\triangleq[\mathbf{e}_{\mathrm{A},1}^{\mathrm{oc}},\mathbf{e}_{\mathrm{A},2}^{\mathrm{oc}},\ldots,\mathbf{e}_{\mathrm{A},N}^{\mathrm{oc}}]\in\mathbb{C}^{2L\times N}\)  and \(\mathbf{E}_{\mathrm{P}}^{\mathrm{oc}}\triangleq[\mathbf{e}_{\mathrm{P},1}^{\mathrm{oc}},\mathbf{e}_{\mathrm{P},2}^{\mathrm{oc}},\ldots,\mathbf{e}_{\mathrm{P},NQ}^{\mathrm{oc}}]\in\mathbb{C}^{2L\times NQ}\)  collect the open-circuit radiation response bases of all antenna and pixel ports, respectively. Each column of $\mathbf{E}_{\mathrm{A}}^{\mathrm{oc}}$ and $\mathbf{E}_{\mathrm{P}}^{\mathrm{oc}}$ contains the \(\theta\)- and \(\phi\)-polarized radiation components over \(L\) spatial (angular) samples.

%\textcolor{blue}{Each column corresponds to the response generated by a unit current at one port with all other ports open-circuited. Since each basis vector contains both $\theta$- and $\phi$-polarized components, the switch configuration $\mathbf b$ reconfigures the array response in both the angular and polarization domains.} 

Unlike prior work, we model the full array and simulate it using CST, as illustrated in Fig.~1, instead of relying on isolated antenna element descriptions. Thus, the MC effect is inherently incorporated in the extracted open-circuit radiation response bases. The overall array radiation response  in~\eqref{eq:pattern_n}  is a linear combination of the individual antenna radiation responses, with the excitation currents serving as the weighting coefficients. Accordingly, the radiation responses of all the antennas in the array are aggregated into the following matrix
\begin{equation}\label{eq:E_b}
\mathbf{E}_{\mathrm{T}}(\mathbf{b}) \triangleq \mathbf{E}_{\mathrm{A}}^{\mathrm{oc}}-\mathbf{E}_{\mathrm{P}}^{\mathrm{oc}}
(\mathbf{Z}_{\mathrm{PP}}+\mathbf{Z}(\mathbf{b}))^{-1}\mathbf{Z}_{\mathrm{PA}}
\in
\mathbb{C}^{2L\times N},
\end{equation}
whose $n$-th column represents the radiation response of the $n$-th antenna element under the given switch-state vector $\mathbf{b}$. 
Therefore, by appropriately designing $\mathbf{b}$, the radiation responses can be flexibly reconfigured to effectively account for mutual coupling effects, enabling MC-aware reshaping of the overall array response to improve system performance.

\section{System Model and Problem Formulation}

We consider a downlink communication system, where \(K\) single-antenna users are served by a BS equipped with a pixel antenna array. 
Based on the above characterization of the pixel array radiation response in \eqref{eq:E_b}, the EM-domain channel from the BS to the $k$-th user is given by
\begin{equation}\label{eq:h}
\mathbf{h}_{\mathrm{EM},k}(\mathbf{b}) = \mathbf{E}_{\mathrm{T}}(\mathbf{b})^{H}\mathbf{H}_{\mathrm{V},k}^{H}\mathbf{e}_{\mathrm{R},k}\in\mathbb{C}^{N},
\end{equation}
where 
$\mathbf{H}_{\mathrm{V},k}\in\mathbb{C}^{2L\times 2L}$ is the virtual channel matrix that represents the channel in the angle-polarization domain. Each entry of $\mathbf{H}_{\mathrm{V},k}$ corresponds to the gain between the polarization component associated with a discretized angle of departure (AoD) and angle of arrival (AoA) grid. The vector $\mathbf{e}_{\mathrm{R},k}$ denotes the radiation response of the receive antenna. 

The transmitted signal is first processed in the baseband domain and subsequently radiated through the pixel antenna array in the EM domain. The received signal at the $k$-th user is given by
\begin{equation}\label{eq:y}
y_{k} = \mathbf{h}_{\mathrm{EM},k}^{H}(\mathbf{b}) \mathbf{w}_{k} s_{k} +\sum_{j\neq k}^{K}\mathbf{h}_{\mathrm{EM},k}^{H}(\mathbf{b}) \mathbf{w}_{j} s_{j} + n_{k},
\end{equation}
where $\mathbf{W}\triangleq[\mathbf{w}_{1}, \mathbf{w}_{2}, \ldots, \mathbf{w}_{K}]\in \mathbb{C}^{N\times K}$ represents the baseband precoder, \(n_{k} \sim \mathcal{CN}(0,\sigma^2)\) is additive white Gaussian noise, and $\mathbf{s}\triangleq[s_{1}, s_{2}, \ldots, s_{K}]^{T}\in \mathbb{C}^{K}$ denotes the transmitted symbols, which are assumed to satisfy $\mathbb{E}\{\mathbf{s}\mathbf{s}^{H}\}=\mathbf{I}_{K}$.

In this paper, we propose to jointly design  the binary switch-state vector $\mathbf{b}$ and the transmit precoder $\mathbf{W}$ to maximize the sum rate under a transmit power constraint. 
The optimization problem can be formulated as 
\begin{subequations}\label{eq:opt_problem}
\begin{align}
\label{eq:opt_problem_a}
\max_{\mathbf{b},\,\mathbf{W}} \quad &
\sum_{k=1}^{K}\log_2\bigg(1+\frac{|\mathbf{h}_{\mathrm{EM},k}^{H}(\mathbf{b})\mathbf{w}_{k}|^2}{\sum_{j\neq k}|\mathbf{h}_{\mathrm{EM},k}^{H}(\mathbf{b})\mathbf{w}_{j}|^2+\sigma^2}\bigg) \\
\mathrm{s.t.} \quad 
\label{eq:opt_problem_b}
&\mathbf{b} \in \{0,1\}^{NQ} , \\
\label{eq:opt_problem_c}
&
\|\mathbf{W}\|_F^2 \leq P_{\mathrm{T}}, 
\end{align}
\end{subequations}
where \(P_{\mathrm{T}}\) denotes the total transmit power.

\section{Joint Switch-State and Precoder Optimization}
In this section, we present the proposed MC-aware EM and signal processing co-design method that jointly optimizes  the binary switch-state $\mathbf{b}$ and the transmit precoder $\mathbf{W}$. 
To effectively tackle the non-convexity of \eqref{eq:opt_problem}, we adopt a heuristic approach based on the SEBO algorithm to optimize the binary switch-state $\mathbf{b}$ for a given precoder $\mathbf{W}$. For a fixed switch state $\mathbf{b}$, the transmit precoder $\mathbf{W}$ is designed based on the FP algorithm. The two sets of variables are updated in an alternating manner until convergence.

\subsection{Antenna Switch-State Design}
With the transmit precoder $\mathbf W$ fixed, the antenna
configuration optimization can be reformulated as
\begin{equation}
\begin{aligned}
\max_{\mathbf b}\quad & f(\mathbf b;\mathbf W)  \\
{\rm s.t.}\quad & \mathbf{b} \in \{0,1\}^{NQ} ,
\end{aligned}\label{eq:b_opt_problem}
\end{equation}
where $f(\mathbf b;\mathbf W)$ denotes the sum rate defined
in (\ref{eq:opt_problem_a}).
%The high-dimensional binary variable $\mathbf{b}$ makes the problem in~\eqref{eq:b_opt_problem} difficult to solve, since an exhaustive search  over all possible switch-state combinations requires $2^{NQ}$ evaluations, which is computationally prohibitive.   To address this issue, we adopt the SEBO algorithm to sequentially optimize the low-dimensional sub-blocks of $\mathbf{b}$, which are then combined to form candidate switch-state configurations with a favorable performance-complexity tradeoff.

The high-dimensional binary variable $\mathbf{b}$ makes the problem
in~\eqref{eq:b_opt_problem} difficult to solve, since a global exhaustive
search over all possible switch-state combinations requires $2^{NQ}$
objective evaluations, which becomes computationally prohibitive as
the number of switches increases. Following the SEBO strategy in~\cite{H.Li_MultiUser},
we optimize the binary switch-state vector in a block-wise manner.
Compared with the global exhaustive search, SEBO sequentially searches
over low-dimensional sub-blocks, substantially reducing the computational
complexity while preserving the discrete nature of the original problem.
Moreover, each block update selects the best configuration with the
remaining variables fixed, resulting in a non-decreasing objective value
and fast convergence in practice.

\begin{algorithm}[!t]
\small
\caption{Joint Switch-State and Precoder Optimization}
\begin{algorithmic}[1]
\label{alg:SEBO}
  \REQUIRE{$\!\!N$, $Q$, $K$, $L$, $U$, $\{\mathbf{H}_{\mathrm{V},k},\mathbf{e}_{\mathrm{R},k}\}_{k=1}^{K}$, $\mathbf{Z}_{\mathrm{AA}}$,  $\mathbf{Z}_{\mathrm{AP}} $,  
    $\mathbf{Z}_{\mathrm{PA}}$, $\mathbf{Z}_{\mathrm{PP}}$, \\
   $\ \ \ \ Z_{\mathrm{on}}$, $Z_{\mathrm{off}}$, $\mathbf{E}_{\mathrm{A}}^{\mathrm{oc}}$, $\mathbf{E}_{\mathrm{P}}^{\mathrm{oc}}$, $P_{\mathrm{T}}$, $\sigma^{2}$.}
    \ENSURE $\mathbf{b}$ and $\mathbf{W}$. 
    \STATE {Initialize $\mathbf{b}$ and $\mathbf{W}$.}
     \WHILE{no convergence of the sum rate}
            \WHILE{no convergence of $f(\mathbf b;\mathbf W)$}
                   \FOR {$u=1:U$}
                        \STATE {Update the $u$-th block  $\mathbf{b}_{\mathcal{I}_u}$ by  \eqref{eq:block_update}.}
                    \ENDFOR
            \ENDWHILE
             \STATE{Construct the EM-domain channel $\mathbf{H}_\mathrm{EM}$ \eqref{eq:h}.}
        \WHILE{no convergence of $f(\mathbf W;\mathbf b)$}
            \STATE {Update  $\bm{\mu}$ and $\bm{\xi}$  by   \eqref{eq:mu_xi}.}
 
            \STATE {Update the transmit precoder $\mathbf{W}$ by   \eqref{eq:W}.}
     \ENDWHILE
     
     \ENDWHILE

     \STATE {Return $\mathbf{b}$ and $\mathbf{W}$.}
\end{algorithmic}
\end{algorithm}
 
The main idea of SEBO is to partition the binary switch-state vector $\mathbf{b}$ into $U$ disjoint blocks and optimize them sequentially. Let $J = \left\lceil {NQ}/{U} \right\rceil$
denote the nominal block length. Then, the indices of the  switches within the $u$-th block are included in  the set $\mathcal{I}_u$, which yields 
\begin{equation}
\!\!\!\!\!\!\mathcal{I}_u\!=\!
\left\{(u-1)J+1,\dots,\min(uJ,NQ)\right\}\!\!,
 ~u\!=\!1,2,\dots,U.\!\!\!\!\!
\end{equation}
The corresponding block size is $J_u = |\mathcal{I}_u|$. 

Starting from an initial switch-state vector $\mathbf{b}$, SEBO updates one block at a time. When optimizing the $u$-th block, all the other blocks are fixed, and only the sub-vector $\mathbf{b}_{\mathcal{I}_u}$ is updated by solving
\begin{equation}
\label{eq:block_update}
\mathbf{b}_{\mathcal{I}_u}
=
\arg\max_{\mathbf{c}\in\{0,1\}^{J_u}}
f\!\left(
\mathbf{c};\mathbf{b}_{\bar{\mathcal{I}}_u},\mathbf{W}
\right),
\end{equation}
where the vector $\mathbf{c}\in\{0,1\}^{J_u}$ represents the optimization variables corresponding to the switch states in the $u$-th block, and  $\bar{\mathcal{I}}_u$ denotes the complementary index set of $\mathcal{I}_u$. After updating $\mathbf{b}_{\mathcal{I}_u}$, this block is fixed, and the algorithm proceeds to optimize the next block in the same manner. This sequential procedure is repeated until all blocks have been updated. Since the block size $J_u$ is much smaller than $NQ$, the optimization in \eqref{eq:block_update} can be carried out by exhaustive search over the reduced search space of cardinality $U2^{J_u}$, thereby achieving a favorable tradeoff between performance and complexity. The transmit precoder $\mathbf{W}$ is kept fixed during the SEBO update and is updated afterward using the FP algorithm.

%\subsubsection{Random perturbation}
%Although the block-wise exhaustive search can effectively improve the objective, it is still a local optimization procedure and may get trapped in poor local optima. To further enhance the search capability, after full block-update sweep, a random perturbation step is introduced.

%Specifically, we randomly select $S$ entries from the binary switch-state vector $\mathbf{b}$, with $S<J$, and flip their values. 
%Let $\mathcal{S}\subseteq \{1, 2, \dots, NQ\}$ denote the selected index set with $|\mathcal{S}|=S$. The perturbed vector $\tilde{\mathbf{b}}$ is constructed as
%\begin{equation}
%\label{eq:b_perturb}
%\tilde b_i
%=
%\begin{cases}
%1-b_i, & i\in\mathcal{S},\\
%b_i, & i\notin\mathcal{S}.
%\end{cases}
%\end{equation}
%The perturbed vector is accepted only if it yields a better objective value, i.e.,
%\begin{equation}
%\label{eq:f_perturb}
%\mathbf{b}
%\leftarrow
%\tilde{\mathbf{b}},\ \ \mathrm{if} \ \ 
%f(\tilde{\mathbf{b}};\mathbf{W}) > f(\mathbf{b};\mathbf{W}).
%\end{equation}
%This perturbation mechanism helps the algorithm escape from unfavorable local stationary points while preserving the monotonic improvement property.

\subsection{Precoder Design}
 
Given the antenna configuration $\mathbf b$, the EM-domain
channels in \eqref{eq:h} are fixed. For notational simplicity,
let $\mathbf h_k\triangleq\mathbf h_{\mathrm{EM},k}(\mathbf b)$.
The fixed-configuration precoder design problem is given by
\begin{equation}
\begin{aligned}
\max_{\mathbf W}\quad &
f(\mathbf W;\mathbf b)
=
\sum_{k=1}^{K}
\log_2\left(1+\gamma_k\right) \\
{\rm s.t.}\quad &
\|\mathbf W\|_F^2 \leq P_{\mathrm T},
\end{aligned}
\label{eq:W_opt_problem}
\end{equation}
where
$
\gamma_k =
\frac{|\mathbf h_k^H\mathbf w_k|^2}
{\sum_{j\neq k}|\mathbf h_k^H\mathbf w_j|^2+\sigma^2}
$
is the SINR of user $k$.
We solve \eqref{eq:W_opt_problem} using the FP method~\cite{M. Liu FP}, where the auxiliary variables
$\bm\mu=[\mu_1,\ldots,\mu_K]^T$ and
$\bm\xi=[\xi_1,\ldots,\xi_K]^T$ are iteratively updated with
the precoder. For a given $\mathbf W$, the auxiliary variables
are updated as
\begin{equation}
\mu_k =
\frac{|\mathbf h_k^H\mathbf w_k|^2}
{\sum_{j\neq k}|\mathbf h_k^H\mathbf w_j|^2+\sigma^2},
\quad
\xi_k =
\frac{\sqrt{1+\mu_k}\mathbf h_k^H\mathbf w_k}
{\sum_{j=1}^{K}|\mathbf h_k^H\mathbf w_j|^2+\sigma^2}.
\label{eq:mu_xi}
\end{equation}
For fixed $\bm\mu$ and $\bm\xi$, the precoder is updated as
\begin{equation}
\mathbf w_k =
\sqrt{1+\mu_k}\xi_k
\left(
\sum_{j=1}^{K}|\xi_j|^2\mathbf h_j\mathbf h_j^H
+ \lambda \mathbf{I}_N \right)^{-1}
\mathbf h_k,\quad \forall k,
\label{eq:W}
\end{equation}
where $\lambda\geq 0$ is the Lagrange multiplier associated with the
transmit-power constraint. For given $\bm{\mu}$ and $\bm{\xi}$, the
transmit power $\|\mathbf W(\lambda)\|_F^2$ is monotonically decreasing
with respect to $\lambda$. Therefore, if the unconstrained solution with
$\lambda=0$ satisfies $\|\mathbf W(0)\|_F^2\leq P_{\mathrm T}$, we set
$\lambda=0$; otherwise, $\lambda>0$ is efficiently determined via a
one-dimensional bisection search such that
$\|\mathbf W(\lambda)\|_F^2=P_{\mathrm T}$. The updates in
\eqref{eq:mu_xi} and \eqref{eq:W} are repeated until convergence.

Following the above derivations, the proposed MC-aware joint
optimization algorithm alternates between SEBO-based switch-state
design and FP-based sum-rate precoder design until convergence,
as summarized in Algorithm~1.

\begin{table}[!t]
\caption{CST Simulation Parameters}
\small
\label{tab:system_settings}
\centering
\renewcommand{\arraystretch}{1.0}
\begin{tabular}{l c c}
\hline 
\textbf{Parameter} & \textbf{Symbol} & \textbf{Value} \\
\hline
Operating frequency                    & $f$                       & $2.4\,\mathrm{GHz}$ \\
Number of antennas           & $N$                       & $4$ \\
Number of switches& $Q$                       & $80$ \\
Number of parasitic pixels    & $N_{\mathrm{pix}}$
      & $50$ \\
Number of ports            & $N+NQ$                  & $324$ \\
Size of pixel antenna           & $D_{\mathrm{pix}}$        & $62.5\,\mathrm{mm}$ \\
Size  of ground           & $D_{\mathrm{gro}}$        & $84\,\mathrm{mm}$ \\
Size  of pixel           & $d_{\mathrm{pix}}$        & $11\,\mathrm{mm}$ \\
Distance between antennas    & $l_{\mathrm{ant}}$                       & $84\,\mathrm{mm}$ \\
Inter-pixel gap    & $l_{\mathrm{pix}}$                       & $1\,\mathrm{mm}$ \\
ON-state impedance          & $Z_{\mathrm{on}}$                  & $10^{-3}\,\Omega$ \\
OFF-state impedance          & $Z_{\mathrm{off}}$                  & $10^{10}\,\Omega$ \\
Number of angle samples            & $L$                  & $72$ \\
Angular resolution            & $\Delta\theta=360^\circ/L$                  & $5^\circ$ \\
 \hline 
\end{tabular}
\end{table}

\section{Simulation Results}
In this section, we provide simulation results to evaluate the effectiveness of the proposed MC-aware EM and signal processing co-design. 
We consider $K=3$ users in a line-of-sight (LoS) propagation environment. In the virtual channel matrix $\mathbf H_{\mathrm V,k}$, the nonzero angle-polarization domain  entries are determined by the LoS directions of the users, and the corresponding complex channel coefficients are scaled by the distance-dependent large-scale channel gain. 
%Specifically, the large-scale power gain is modeled as $10^{-\frac{C_{0}}{10}}(\frac{r_{k}}{D_{0}})^{-\alpha}$ with $C_{0}=30$ dB, $D_{0}=1$m, $\alpha=2.2$, and the user distance $r_{k}\in[50\mathrm{m},100\mathrm{m}]$. 
The noise power is set as $\sigma^2=-70\ \mathrm{dBm}$. The pixel antenna topology adopted in this work follows the design in \cite{X.Zheng_Polarization}. Each antenna consists of a feeding structure and reconfigurable parasitic pixel surfaces implemented on dielectric substrates. The two parasitic layers comprise a total of 50 metallic pixels interconnected by 80 PIN diodes. The substrate and layer configuration follow \cite{X.Zheng_Polarization}, while the main geometric and switching parameters used in our CST model are summarized in Table I.
%The CST simulation parameters are summarized in Table I.
Based on these parameters, full-wave EM simulations are conducted in CST to extract the network and radiation parameters required by the multi-port model in Section II. Specifically, the impedance matrix in \eqref{eq:Z_matrix} and the open-circuit embedded radiation response bases $\mathbf{E}_{\mathrm{A}}^{\mathrm{oc}}$ and $\mathbf{E}_{\mathrm{P}}^{\mathrm{oc}}$ in \eqref{eq:pattern_n} are obtained from the full-array CST model. For any given switch-state vector $\mathbf b$, these extracted parameters are substituted into \eqref{eq:E_b} to construct the switch-state-dependent transmit radiation matrix $\mathbf E_{\mathrm T}(\mathbf b)$. At the user side, an omnidirectional receive antenna is assumed, i.e., $\mathbf e_{\mathrm R,k}$ is fixed and normalized. Combining $\mathbf E_{\mathrm T}(\mathbf b)$, $\mathbf e_{\mathrm R,k}$, and the angle-domain virtual channel $\mathbf H_{\mathrm V,k}$, the user-specific EM-domain channel $\mathbf h_{\mathrm{EM},k}(\mathbf b)$ is obtained according to \eqref{eq:h}. This channel construction is then used for the subsequent antenna configuration and precoder design.
% The CST simulation parameters are summarized in TABLE I. Following the methodology described in Section II, the network parameters are extracted from CST-based full-wave EM simulations to construct the EM-domain channel for the pixel antenna array. These include the impedance parameters in \eqref{eq:Z_matrix} and the open-circuit radiation pattern matrix $\mathbf{E}_{\mathrm{A}}^{\mathrm{oc}},\mathbf{E}_{\mathrm{P}}^{\mathrm{oc}}$ in \eqref{eq:pattern_n}. 

%Based on the constructed channels, we evaluate the proposed MC-aware pixel antenna array design, denoted as ``\textbf{MC-aware, Pixel}'', which uses the full coupled-array model for both design and performance evaluation. As a mismatch baseline, we consider an MC-unaware pixel design, denoted as ``\textbf{MC-unaware, Pixel}'', where inter-element MC is ignored during design but the performance is evaluated using the actual coupled-array response. We also include an idealized MC-free pixel array, denoted as ``\textbf{MC-free, Pixel}'', in which inter-element MC is removed in both design and evaluation. Finally, a conventional patch antenna array with the same aperture is adopted as a fixed-radiation benchmark, with its coupled and idealized MC-free cases denoted as ``\textbf{MC-unaware, Patch}'' and ``\textbf{MC-free, Patch}'', respectively.

We compare the following methods: \textbf{MC-aware, Pixel} uses the full coupled-array model for both switch-state/precoder design and performance evaluation; \textbf{MC-unaware, Pixel} designs the switch states and precoder using a decoupled model that neglects inter-element MC, but evaluates the resulting transmission performance using the actual coupled-array response; \textbf{MC-free, Pixel} is an idealized reference in which inter-element MC is removed in both design and evaluation. For comparison with a fixed-radiation architecture, a conventional patch antenna array with the same aperture is also considered. The corresponding idealized and mismatched patch-array baselines are denoted by \textbf{MC-free, Patch} and \textbf{MC-unaware, Patch}, respectively. %The same total transmit power and the same linear MMSE precoder normalization are used for all schemes.

\begin{figure}[!t]
  \centering
  \includegraphics[width=0.7\linewidth]{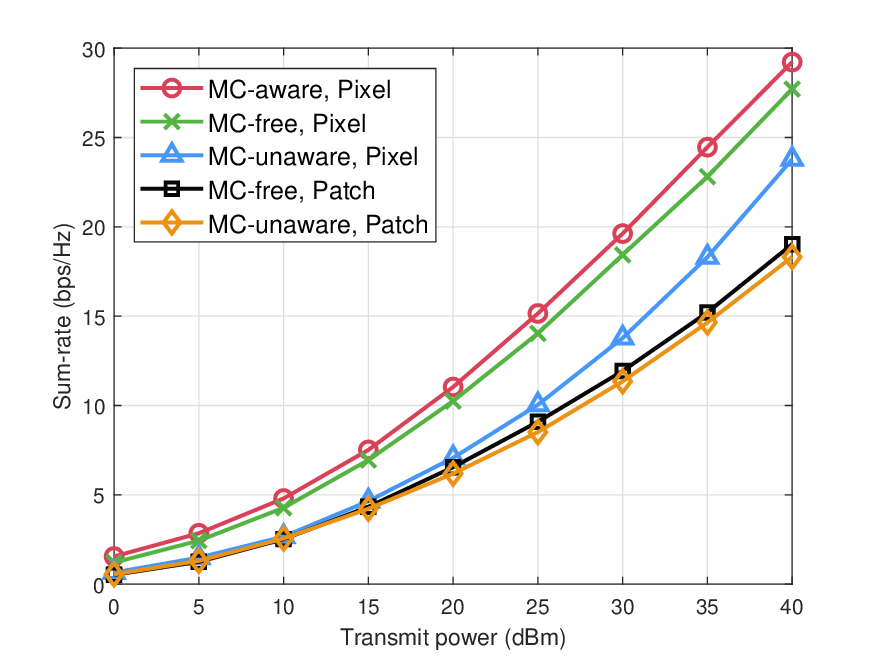}
  \caption{Sum rate versus transmit power $P_{\mathrm{T}}$.}
  \label{fig:sumrate} \vspace{-0.6 cm}
\end{figure}

\begin{figure}[!t]
  \centering
  \includegraphics[width=0.7\linewidth]{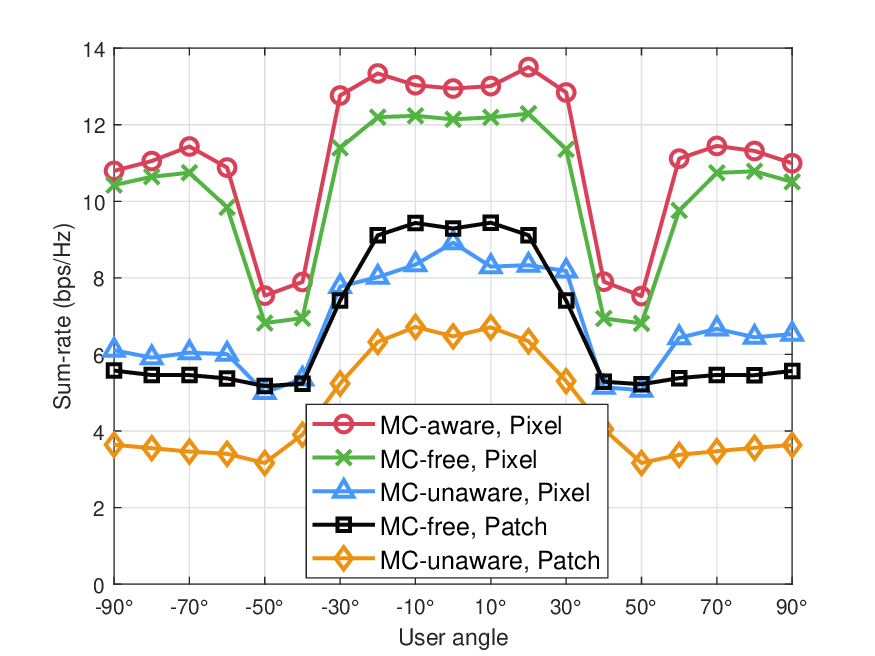}
  \caption{Sum rate versus the AoD of the swept user, where the other two users are fixed
at $\pm45^\circ$, $P_{\rm T} = 20$ dBm. }
  \label{fig:angle_user} \vspace{-0.4 cm}
\end{figure}

\begin{figure*}[!t]
  \centering
  \includegraphics[width=0.75\linewidth]{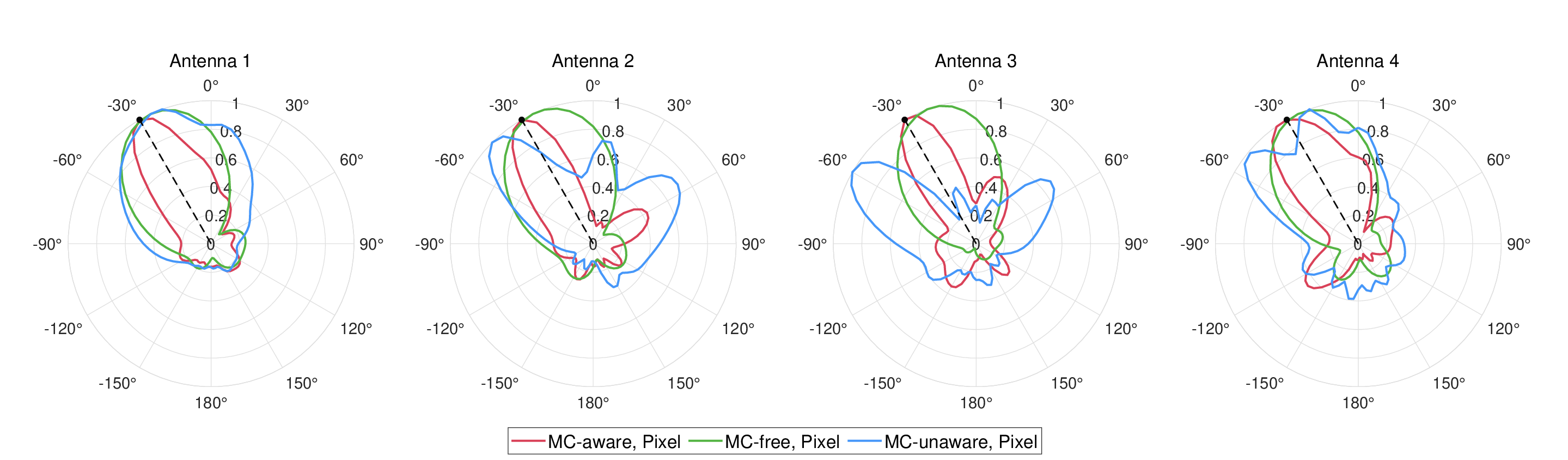}
  \caption{Normalized embedded radiation patterns of the four
pixel antennas for a representative user direction of $-30^\circ$.} \vspace{-0.5cm}
  \label{fig:pattern}
\end{figure*}

Fig.~\ref{fig:sumrate} shows the average sum rate as a function of the transmit power, where the user angles are randomly selected from $[-90^\circ,90^\circ]$. The proposed MC-aware pixel design achieves the highest sum rate over the entire SNR range by reconfiguring both the radiation pattern and polarization response of the array. Its gain over the MC-unaware pixel scheme becomes more evident at medium and high SNRs, indicating that the dominant loss of the MC-unaware design comes from the mismatch between the decoupled design model and the actual coupled-array response, rather than from thermal noise. The MC-aware design also outperforms the MC-free pixel reference under the same aperture and power normalization, demonstrating that mutual coupling is not necessarily an impairment. When accurately modeled, inter-element MC provides additional embedded-response diversity that can be exploited for EM-domain channel shaping. The patch baselines achieve lower rates because their fixed radiation responses provide limited adaptability to the multi-user channels. In contrast, the pixel array introduces additional EM-domain degrees of freedom through the switch-controlled current distribution, allowing its embedded radiation patterns and polarization responses to be jointly adapted with the digital precoder. Therefore, even under the same physical aperture, the pixel architecture can better shape the effective multi-user channels and reduce unfavorable channel correlations, which explains the rate gain over the conventional patch array.

Fig.~\ref{fig:angle_user} evaluates the sum rate as the AoD of one user is swept from $-90^\circ$ to $90^\circ$, while the other two users are fixed at $\pm45^\circ$. The proposed MC-aware pixel design achieves the best overall performance across the considered angular range, showing the effectiveness of jointly optimizing the switch-state-dependent coupled radiation responses and the digital precoder. The rate dips around $\pm45^\circ$ because the swept user becomes angularly close to one of the fixed users, which increases the effective channel correlation and makes spatial interference suppression more difficult.  Even in these challenging regions, the proposed approach outperforms the MC-unaware pixel design, confirming the importance of accounting for the actual coupled-array response during optimization. Its gain over the MC-free pixel reference further suggests that mutual coupling, when accurately modeled, can be exploited as useful embedded-response diversity rather than being treated solely as an impairment. The patch-array baselines exhibit larger degradation due to their limited radiation reconfigurability. 

%Fig.~\ref{fig:angle_user} shows that the proposed MC-aware pixel scheme provides the highest sum-rate across the whole swept-angle range. The rate drops around $\pm45^\circ$ because the swept user becomes angularly close to one of the two fixed users, making spatial multiplexing more difficult. Nevertheless, the proposed method still maintains the best performance in these challenging regions, demonstrating its robustness to unfavorable user-angle configurations. Moreover, its consistent gain over the MC-unaware pixel design confirms the importance of accounting for mutual coupling during optimization, while its advantage over the MC-free pixel benchmark indicates that the properly exploited coupling can provide useful embedded-pattern diversity. The patch baselines degrade more significantly, especially toward unfavorable angular regions, due to their lack of hardware-level radiation reconfigurability.

Fig.~\ref{fig:pattern} illustrates the normalized embedded radiation patterns
of the four pixel antennas for a representative user direction of $-30^\circ$. The three pixel designs lead to clearly different patterns, confirming that inter-element MC changes the mapping from switch states to realized radiation responses. For the MC-aware design, the high-response regions of the embedded patterns are more consistently oriented toward the
desired direction, whereas the MC-free and MC-unaware
designs exhibit less compatible element-wise responses under
the actual coupled-array behavior. This pattern-level observation provides an EM-domain explanation for the sum-rate gain and angular robustness observed in Figs.~\ref{fig:sumrate} and~\ref{fig:angle_user}.

%This pattern-level result explains the gains observed in Figs. 3 and 4: the proposed method optimizes the switch states according to the realized coupled responses, thereby improving EM-domain channel shaping. Since the patterns are normalized, Fig. 5 should be interpreted as qualitative evidence of angular selectivity rather than a direct comparison of absolute realized gain.

\section{Conclusions}
We developed an MC-aware EM and signal-processing co-design framework for downlink MU-MISO systems with pixel antenna arrays. A full-array multi-port circuit model was combined with CST-extracted impedance and
embedded far-field responses to construct a switch-state-dependent EM-domain channel that captured both intra-antenna pixel coupling and inter-element MC. Based on this model, the sum rate was maximized by designing the binary switch-state vector using SEBO and the transmit
precoder via FP. Simulation results showed that the proposed MC-aware design achieved higher sum rates than the MC-unaware, MC-free, and patch-array baselines. These results indicated that inter-element MC
should be explicitly modeled and could be leveraged as part of the EM-domain channel-shaping capability of practical pixel antenna arrays.


\begin{thebibliography}{99}

%%%%%%%%%%%%%%%%%%%% Reconfigurable pixel antennas %%%%%%%%%%%%%%%%%%%
\bibitem{M.Liu_Reconfigurable} M. Liu, M. Li, R. Liu, Q. Liu, and A. L. Swindlehurst, ``Reconfigurable antenna arrays: Bridging electromagnetics and signal processing,'' \textit{IEEE Commun. Mag.}, early access, Apr. 2026, doi:10.1109/MCOM.001.2500651. 

%%%%%%%%%%%%%%%%%%%% ISAC security problem %%%%%%%%%%%%%%%%%%%

\bibitem{S. Shen_Reconfigurable} S. Shen, K.-K. Wong, and R. Murch, ``Antenna coding empowered by pixel antennas,'' \textit{IEEE Trans. Commun.}, vol. 74, pp. 446-460, 2026.


\bibitem{A.Grau_Multifunctional} A. Grau Besoli and F. De Flaviis, ``A multifunctional reconfigurable pixeled antenna using MEMS technology on printed circuit board,'' \textit{IEEE Trans. Antennas Propag.}, vol. 59, no. 12, pp. 4413-4424, Dec. 2011.

\bibitem{D.Rodrigo_Reconfigurable} D. Rodrigo, B. A. Cetiner, and L. Jofre, ``Frequency, radiation pattern and polarization reconfigurable antenna using a parasitic pixel layer,'' \textit{IEEE Trans. Antennas Propag.}, vol. 62, no. 6, pp. 3422-3427, Jun. 2014.

\bibitem{S.Song_Efficient} S. Song and R. D. Murch, ``An efficient approach for optimizing frequency reconfigurable pixel antennas using genetic algorithms,'' \textit{IEEE Trans. Antennas Propag.}, vol. 62, no. 2, pp. 609-620, Feb. 2014.

%%%%%%%%%%%%%%%%%%%% ISAC security solution %%%%%%%%%%%%%%%%%%%

\bibitem{X.Chen_Mutual} X. Chen, S. Zhang, and Q. Li, ``A review of mutual coupling in MIMO systems,'' \textit{IEEE Access}, vol. 6, pp. 24706-24719, 2018.

\bibitem{H.Yang_Mutual} H. Yang, Y. Zhao, K.-K. Wong, H.-H. Chen, and C.-B. Chae, ``Bridging theory and practice in reconfigurable fluid antenna systems,'' \textit{IEEE Wireless Commun.},  vol. 33, no. 4, pp. 109-116, Aug. 2026. 


\bibitem{H.Li_Mutual} H. Li and S. Shen, ``Antenna coding design based on pixel antennas for multi-user MISO systems,'' in \textit{Proc. IEEE 26th Int. Workshop Signal Process. Adv. Wireless Commun. (SPAWC)}, Surrey, United Kingdom, 2025, pp. 1-5.
%%%%%%%%%%%%%%%%%%%% solution limitation %%%%%%%%%%%%%%%%%%%


\bibitem{B.Zuo_Antenna} B. Zuo, S. Shen, and H. Li, ``Antenna coding optimization for pixel antenna empowered wireless communication using deep learning with heterogeneous multi-head selection,'' in \textit{Proc. IEEE Wireless Commun. Netw. Conf. Workshops (WCNCW)}, Kuala Lumpur, Malaysia, 2026, pp. 1-6.

%%%%%%%%%%%%%%%%%%%% reconfigurable antenna solution %%%%%%%%%%%%%%%%%%%
\bibitem{H.Li_MultiUser} H. Li and S. Shen, ``Antenna coding design for multi-user transmissions using pixel antennas,'' \textit{IEEE Trans. Cog. Commun. and Networking}, vol. 12, pp. 7634-7646, 2026.

\bibitem{A.Sams_Optimizing} A. Sams, Y.-C. Hsiao, M. Talha, B. Smida, and A. Sabharwal, ``Optimizing ISAC MIMO systems with reconfigurable pixel antennas,'' in \textit{Proc. 59th Asilomar Conf. Signals, Syst., Comput.}, Pacific Grove, USA, 2025, pp. 896-901.

\bibitem{benitzhak_cst}
G. Ben-Itzhak, M. Saavedra-Melo, B. Bradshaw, E. Ayanoglu, F. Capolino, and A. L. Swindlehurst, ``Design and operation principles of a wave-controlled reconfigurable intelligent surface,'' \textit{IEEE Open J. Commun. Soc.}, vol. 5, pp. 7730-7751, 2024.

\bibitem{M. Liu FP} M. Liu, M. Li, R. Liu and Q. Liu, ``Tri-hybrid beamforming design with reconfigurable antennas,'' \textit{IEEE Trans. Wireless Commun.}, to appear.

\bibitem{X.Zheng_Polarization}
W. Zheng, Y. Yang, and H. Li, ``Design of polarization-reconfigurable pixel antennas with optimized PIN-diode implementation,'' \textit{IEEE Trans. Antennas Propag.}, vol. 73, no. 2, pp. 851-862, Feb. 2025.


\end{thebibliography}
\end{document}